\documentclass[preprint,aps,showpacs]{revtex4}

\usepackage{graphicx}
\usepackage{dcolumn}
\usepackage{bm}

\nonstopmode
\usepackage{epsfig}
\graphicspath{{./}{data_paper/}}
\DeclareGraphicsExtensions{.ps,.ps.gz}
\begin{document}

\title{\bf Studying the single--electron transistor by photoionization}

\author{Ioan B\^aldea}
\email{ioan@pci.uni-heidelberg.de}
\altaffiliation[Also at ]{National Institute for Lasers, Plasmas, and Radiation Physics, 
Institute for Space Sciences, RO 077125 Bucharest, Romania.}
\author{Horst K\"oppel} 
\affiliation{Theoretische Chemie,
Physikalisch-Chemisches Institut, Universit\"{a}t Heidelberg, Im
Neuenheimer Feld 229, D-69120 Heidelberg, Germany}

\date{\today}

\begin{abstract}
We report theoretical results demonstrating that 
photoionization can be a useful tool for investigating  
single--electron transistors. It permits to obtain information 
on the quantum dot occupancy and the charging energy in a 
\emph{direct} manner, and not indirectly, as done in transport experiments. 
It is worth emphasizing that 
in the photoionization processes considered by us, an electron
absorbs a photon with energy of the order of the work functions
($\sim 1$\,eV) and is ejected into the vacuum. 
This phenomenon is completely different from
the widely investigated photo-assisted tunneling. There, an electron 
tunnels through a Coulomb island from one electrode to another
by absorbing a photon of much lower energy, 
of the order of the charging energy
(typically, a few meV).
Suggestions are given on how to conduct experiments 
using photoionization alone 
or in combination with transport measurements. 
Monitoring zero kinetic energy (ZEKE) photoelectrons is
especially recommended, because ZEKE--spectroscopy offers 
a better resolution than standard photoemission.
\end{abstract}

%
%
%
%

%
%
%
%

%
%
%
%

%

\maketitle

\renewcommand{\topfraction}{1}
\renewcommand{\bottomfraction}{1}
\renewcommand{\textfraction}{0}
\section{Introduction}
\label{sec-introduction} 
In a single--electron transistor (SET), a quantum dot (QD) is coupled 
via tunneling junctions to metallic electrodes \cite{FultonDolan:87,Goldhaber-GordonNature:98}.
An isolated quantum dot (QD) behaves like an ``artificial'' atom, 
where the electron motion is confined within a small region with size of the 
order of nanometers \cite{Kastner:93,KouwenhovenMarcus:98}.  
Similar to atoms, the single electron levels of a QD are well separated 
energetically.
Experiments on SET transport are often interpreted within the Anderson 
impurity model at equilibrium \cite{AverinLikharev:86,Ng:88,Glazman:88,Izumida:98}. 
The QD is modeled by a single localized state. Its energy $\varepsilon_d$ 
can easily be tuned by varying the potential $V_g$ of a gate.
This approximation is 
particularly justified in small QDs, where 
single electron levels are well separated energetically.  
This state can accommodate 
$n_d=0, 1, \mbox{ or } 2$ electrons.
A key quantity for electronic transport through the SET is just 
this number $n_d$ of (``valence'') electrons.
\par
Unlike in extended systems, a 
nonvanishing charging energy $U=e^2/2C$ has to be paid to add 
one electron to a QD ($C$ is the total capacitance). At sufficiently low 
temperatures, for a small QD 
weakly coupled to electrodes, the charging energy $U$ represents the dominant 
energy scale; it exceeds the thermal energy and the finite width 
$\Gamma $ of the dot level resulting from its 
hybridization ($t_d$) with electron states of the electrodes. 
\par
Therefore, charge fluctuations are largely suppressed and the number 
of electrons of the dot is an integer 
in broad $\varepsilon_d$-ranges (plateaus): 
$n_d \simeq 0$ for $\varepsilon_d > \varepsilon_F + \Gamma $,
$n_d \simeq 1$ for 
$\varepsilon_F - U + \Gamma < \varepsilon_d < \varepsilon_F - \Gamma $, and 
$n_d \simeq 2$ for $\varepsilon_d < \varepsilon_F - U - \Gamma$.
The transitions between the 
states with well defined dot charge occur within narrow 
$\varepsilon_d$-ranges of width $\sim 2\Gamma$. In these ranges, 
the QD state is a combination of two nearly degenerate states of well 
defined charge (mixed valence regimes). There, the dot charge can 
fluctuate and, thus, electron transport becomes possible.
This gives rise to the well known Coulomb blockade peaks, separated by 
$\Delta \varepsilon_d \simeq U$ in the curve of the conductance $G(\varepsilon_d)$
at low temperatures (but larger than the Kondo temperature $T_K$) 
\cite{KouwenhovenMarcus:98}.
Although charge fluctuations are suppressed within the plateaus, 
spin fluctuations 
through virtual intermediate state are allowed 
within the (Kondo) plateau where the QD is occupied by one electron 
($n_d \simeq 1$).
At $T < T_K$, they yield a narrow peak (Kondo resonance) in the QD density of states, 
which enhances the conductance up to the 
value of the ideal point contact $G_0=2e^2/h$ (unitary limit) \cite{Wiel:00}.
\par
The experimental observation of the Coulomb blockade phenomenon and the Kondo 
effect in electric transport of SETs is remarkable \cite{Goldhaber-GordonPRL:98,Wiel:00}. 
It \emph{indirectly} confirms the occurrence of the charge plateaus in broad $V_g$--ranges. 
This evidence is indirect, because what one directly measures in experiment 
is the conductance $G$, and not the dot charge $n_d$. Besides, the charging energy $U$ is also 
indirectly estimated as $\Delta \varepsilon_d \simeq U$, while the 
experimentally measured quantity is $V_g$ and not $\varepsilon_d$.
\par
In the present paper, we shall propose an alternative method to investigate 
phenomena related to 
the charge plateaus in a SET, namely the photoionization. 
Photoionization is known as a very useful tool to study strong electron correlations 
\cite{Cederbaum:1986,Baldea:2002}. As we shall show, from the investigation 
of the ionization one can extract more direct information on the QD charge 
than from transport experiments.
\par
The remaining part of the paper is organized as follows. In Section \ref{sec-results}, 
we describe the model and present the theoretical results on photoionization. 
Based on these theoretical results, in Section \ref{sec-exp} we suggest possible 
experiments to employ photoionization as a tool for investigating SETs. 
Conclusions are presented in the final Section \ref{sec-conclusion}.
\section{Model and theoretical results}
\label{sec-results} 
We shall also describe the SET by the Anderson model
\begin{eqnarray}  
H & = & 
\varepsilon_{F} \sum_{\sigma, l=-1}^{-M_L} a_{l,\sigma}^{\dagger} a_{l,\sigma}^{}
+ \varepsilon_{F} \sum_{\sigma, l=1}^{M_R}  a_{l,\sigma}^{\dagger} a_{l,\sigma}^{}
\nonumber \\ 
& &
- t \sum_{\sigma,l=-1}^{-M_L+1} \left(
  a_{l,\sigma}^{\dagger} a_{l-1,\sigma}^{} 
+ h.c. \right) \nonumber \\
& &
-t \sum_{\sigma,l=1}^{M_R-1} \left(
  a_{l,\sigma}^{\dagger} a_{l+1,\sigma}^{} 
+ h.c. \right) \label{eq-hamiltonian} \\
& & 
- t_{d} \sum_{\sigma} \left(a_{-1,\sigma}^{\dagger} d_{\sigma}^{} +  a_{+1,\sigma}^{\dagger} d_{\sigma}^{} 
+ h.c. \right) 
\nonumber \\
& &
+ \varepsilon_{d}  \sum_{\sigma} d_{\sigma}^{\dagger} d_{\sigma}^{}
+  U \hat{n}_{d,\uparrow}^{} \hat{n}_{d,\downarrow}^{}, \nonumber
\end{eqnarray}  
where $a_{l,\sigma}$ ($a^{\dagger}_{l,\sigma}$) denote 
annihilation (creation) operators for electrons of spin $\sigma$ in the left and right 
leads ($L, R$),  
$d_{\sigma} \equiv a_{0,\sigma}$ ($d^{\dagger}_{\sigma}  \equiv a^{\dagger}_{0,\sigma}$) 
destroys (creates) electrons in the QD, and 
$\hat{n}_{d,\sigma} \equiv d_{\sigma}^{\dagger} d_{\sigma}^{}$
are electron occupancies per spin direction.
The QD-electrode coupling is characterized by 
the hopping integral $t_d$, which can be experimentally controlled. 
In the experimental setup of Ref.\  \onlinecite{Goldhaber-GordonPRL:98}, $t_d$ 
can be changed by varying the potential of the gates that form the constrictions,
and the dot energy $\varepsilon_d$ can be tuned by varying the potential of 
a ``plunger'' gate electrode $V_g$
\begin{equation}
\varepsilon_d = \alpha V_g + const .
\label{eq-epsilon_g}
\end{equation}
The number of electrons will be assumed 
equal to the number of sites ($N=M_L + M_R + 1$).
In the photoionization of the QD, 
a photon of energy $\omega$ is absorbed by an electron, 
which is ejected from the QD into vacuum, and brings the ionized system in one of its eigenstates  
$\Psi_k $.
The ionization process can be characterized by an energy threshold 
$\omega_{k}$=$\langle \Psi_k \vert H \vert \Psi_k \rangle$-$ 
\langle \Phi \vert H \vert \Phi \rangle $ 
and a spectroscopic factor $f_{k,\sigma}$
\begin{equation}
\displaystyle
f_{k,\sigma} = 
\vert\langle \Psi_{k}\vert d_{\sigma}\vert 
\Phi \rangle\vert^{2} . 
\label{eq-w}
\end{equation} 
Here $\Phi$ stands for the neutral ground state 
(case $T=0$). The spectroscopic factor is 
directly related to the weight of a line in the ionization spectrum, more precisely 
to the partial--channel ionization cross section \cite{Cederbaum:1986}.
From Eq.\ (\ref{eq-w}) one can deduce the following sum rule
\begin{equation}
\displaystyle
\sum_{k} f_{k,\sigma} = 
\langle \Phi \vert d_{\sigma}^{\dagger}  d_{\sigma} \vert \Phi\rangle \equiv n_{d,\sigma}. 
\label{eq-sum-rule}
\end{equation} 
Eq.\ (\ref{eq-sum-rule}) represents an important result. It permits to 
directly relate the integrated weight of the ionization spectrum 
to the number of electrons on the QD in the neutral ground state.
From a technical side, Eq.\ (\ref{eq-sum-rule}) turns out to be very 
useful to test the numerical results.
It allows to check whether the Lanczos algorithm targets all the 
relevant ionized states $\Psi_k$.
\par
To compute the ionization spectrum, we shall employ the method of exact 
numerical (Lanczos) diagonalization. Although this method can only be applied 
to small electrodes, the results are relevant provided that the number 
of sites of the electrodes are properly chosen to mimic a metallic behavior 
\cite{Baldea:2008b}. 
The single-particle energies of an isolated electrode 
lie symmetrically around $\varepsilon_F$ within the range 
$\left( \varepsilon_F - 2 t, \varepsilon_F + 2 t\right)$.
To ensure that in the ground state of the isolated electrodes 
the Fermi level is 
occupied by one electron, one should consider that each electrode 
consists of an odd number ($M_{L,R}$) of sites. 
To demonstrate the reduced role of the finite-size effects, 
we have also considered the QD embedded in a ring 
($a_{-M_L, \sigma} \equiv a_{M_R, \sigma}$). 
In this case, for a similar reason, the ring without the QD should 
consist of an odd number of sites (odd $M_L + M_R$).
In all numerical results presented here, we set $t=1$. For the numerical calculations, the value 
of the Fermi energy in electrodes $\varepsilon_F$ is \emph{not} important: 
it only fixes the energy zero. Therefore, we do not fix the value of $\varepsilon_F$ 
and present in all the figures results for energies relative to $\varepsilon_F$.
\par
\begin{figure}[htb]
\centerline{\hspace*{-0ex}
\includegraphics[width=0.3\textwidth,angle=-90]{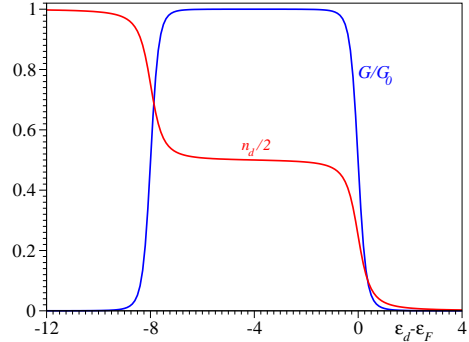}
}
\caption{(Color online) Results for the dot occupancy $n_d$ and normalized 
conductance $G/G_0$ for $U=8$, $t_d=0.2$, $t=1$, 
and $M_L = M_R = 5$ ($N=11$). These curves can be hardly distinguished 
within the drawing 
accuracy from those for open chains with $(M_L, M_R)$=$(3, 3)$, $(3, 5)$, and $(3, 7)$, 
as well as for rings with $N=6, 10$.} 
\label{fig_n_d_G}
\end{figure}

Our numerical results for $n_d$ of Fig.\ \ref{fig_n_d_G} 
demonstrate the formation of broad, well defined charge plateaus 
($n_d \approx 0, 1, 2$), separated by narrow mixed valence regions, 
as discussed above. 
The fact that the $n_d$--curve (as well as those for the bright diabatic 
states, see below) for various electrode sizes and boundary 
conditions are almost identical within the drawing accuracy in 
Fig.\ \ref{fig_n_d_G} demonstrates that finite-size effects are not 
important and the study of small systems is physically relevant.
Our extensive numerical results demonstrate that 
what is essential for the weak finite-size effects is also essential 
for the occurrence of the charge plateaus, namely a width parameter 
$\Gamma \propto t_d^2/t$ smaller than the charging energy $U$. 
On this basis, we argue that 
the charge and the ionization spectra of a QD connected to semi--infinite leads  
can be accurately deduced by computations for such small ``metallic'' electrodes.
\par
Although it does not represent a central issue from the present perspective, 
the weakness of finite-size effects is also relevant for the Kondo effect, which 
we only mention here in passing.
As is well known, essential for the occurrence of this effect is the 
Kondo screening cloud. The latter extends over a 
number of sites $\xi_{K} \sim t/T_{K}$, which rapidly grows (exponentially 
for large $U$) far beyond electrode sizes treatable by exact numerical 
diagonalization. Therefore, it is impossible to obtain the conductance within 
standard transport approaches by employing short electrodes. 
For semi-infinite electrodes, 
the zero-bias conductance $G$ can be computed via the Friedel-Langreth sum rule 
\cite{Langreth:66,Shiba:75,Meir:92}
\begin{equation}  
G = G_0 \sin^2(\pi n_{d}^{\infty}/2)
\label{eq-friedel}  
\end{equation}  
once the dot occupancy $n_{d}^{\infty}$ is known.
Eq.\ (\ref{eq-friedel}) was deduced for a QD attached to semi-infinite leads with a 
continuum density of states, which is the real nanosystem of interest.
The pleasant thing is that the zero-bias conductance $G$ for this case can be solely
determined from $n_{d}^{\infty}$, and a reliable estimated of the latter suffices 
for this purpose. Fortunately,
to this we need not to carry out calculations for semi-infinite or very large 
electrodes: in view of the weak finite-size effect mentioned above, 
$n_d$ computed for short electrodes does represent an accurate estimate, 
$n_{d}^{\infty} \simeq n_d$. 
Indeed, the curve for $G(\varepsilon_d)$ obtained by inserting the calculated 
$n_d$ calculated for short electrodes instead od $n_{d}^{\infty}$, 
also presented in Fig.\ \ref{fig_n_d_G},
has a similar appearance to that obtained by means of the numerical 
renormalization group \cite{Izumida:01,Costi:01} for semi-infinite electrodes and 
nicely reveals the occurrence of the Kondo plateau.
\par
Let us now switch to the results for SET photoionization. To avoid misunderstandings, 
we emphasize that the photoionization considered in this paper is the 
\emph{qualitatively} 
different from the widely-studied photo-assisted tunneling (see, for example,
Ref.\ \onlinecite{Mii:01} and references therein).
The process where a photon helps an electron to tunnel from one electrode to another 
through a Coulomb island might also be viewed as an ``ionization'' phenomenon. 
Nevertheless, the scale of the corresponding ``ionization'' 
energies is determined by the charging energy $U$. 
For semiconducting QDs, it is of the order of $\sim 1$\,meV, that is, much smaller 
than ionization energies of the order of the work functions ($\sim 1$\,eV) considered by us. 
\par
\begin{figure}[htb]
\centerline{\hspace*{-0ex}
\includegraphics[width=0.3\textwidth,angle=-90]{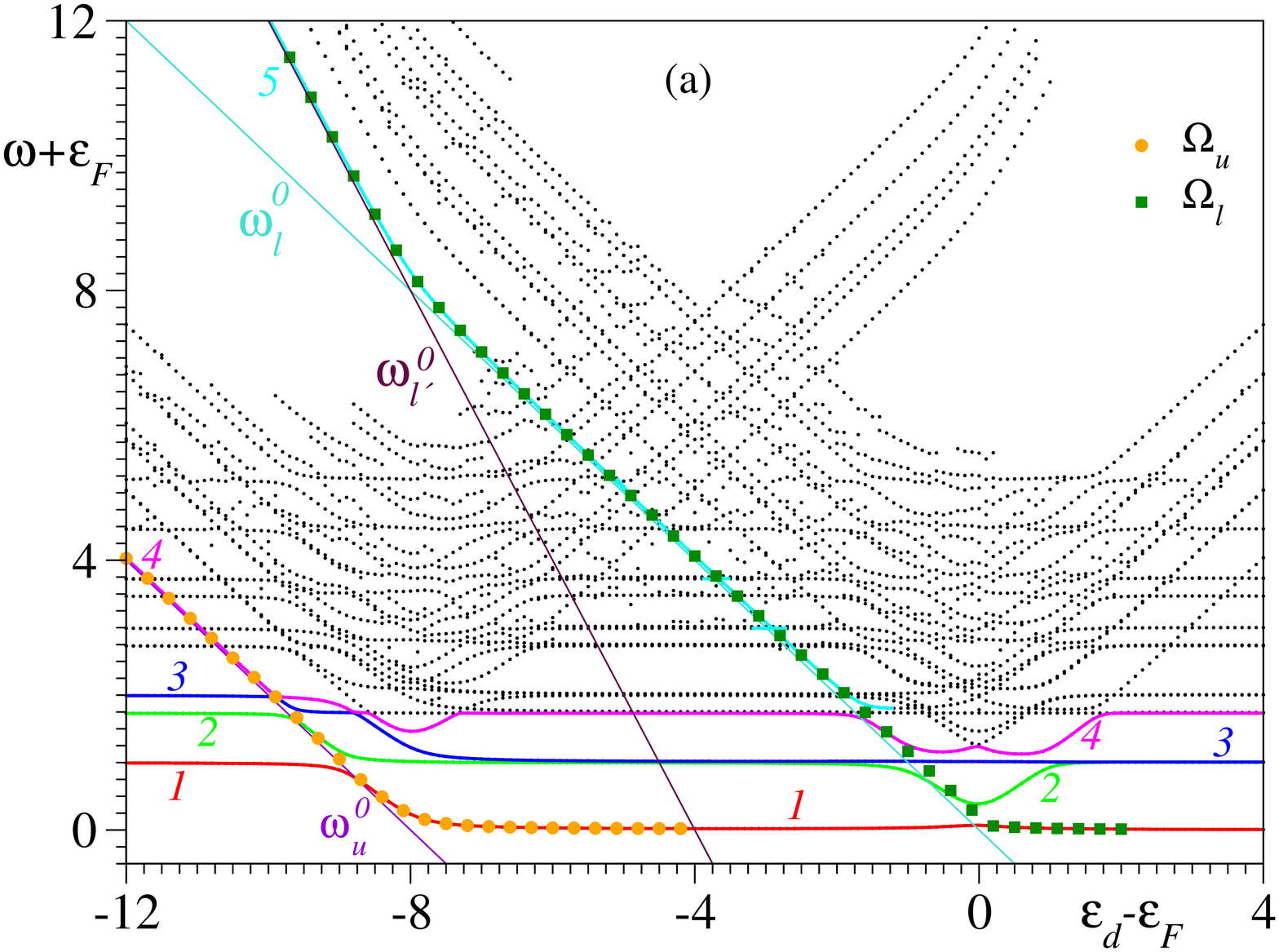}}
\centerline{\hspace*{-0ex}
\includegraphics[width=0.3\textwidth,angle=-90]{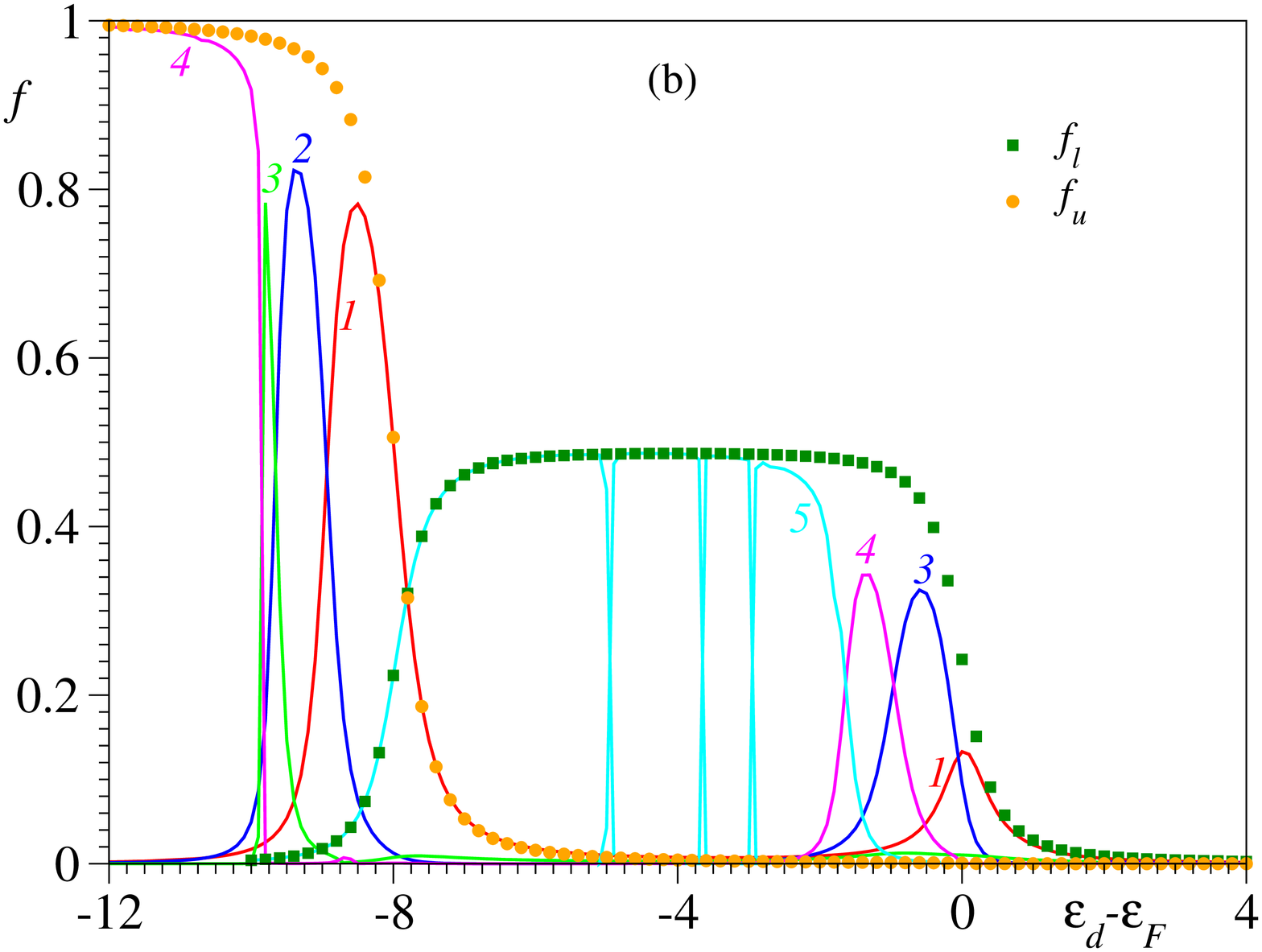}}
\caption{(Color online) Results for ionization energies (a) and spectroscopic factors (b). 
The solid lines 1, 2, 3, and 4 are for the eigenstates with substantial 
spectroscopic factors. The results for the two relevant diabatic bright states 
(\emph{u, l}) are depicted by symbols. 
The black points in (a), which correspond to the ionized eigenstates with 
small spectroscopic factors ($10^{-5} \alt f \alt 10^{-2}$), are not quantitatively significant 
for the ionization spectrum, but are showed to better visualize the occurrence of the 
avoided crossings. Notice that the ionization energies are measured relative to the 
electrode work function ($-\varepsilon_F$), a quantity much larger than the charging energy $U$. 
Parameter values as in Fig.\ \ref{fig_n_d_G}. 
See the main text for details.} 
\label{fig_IP}
\end{figure}

Typical exact numerical results for the ionization 
energies and spectroscopic factors of all the significant ionization processes 
are depicted in Fig.\ \ref{fig_IP}.
To understand these results it is helpful to analyze first the 
limit $t_d \to 0$.
\par
In this limit, two ionization processes of the QD are possible. 
For $\varepsilon_d < \varepsilon_F - U $, the dot is doubly occupied ($n_d = 2$), 
and the 
relevant ionization process consists of removing an electron from the upper 
Hubbard ``band''. The corresponding ionization energy is 
$\omega_{u}^{0} = -\varepsilon_d - U $. 
For $ \varepsilon_F - U < \varepsilon_d < \varepsilon_F $, the dot is 
occupied by a single electron in the lower Hubbard ``band'' ($n_d = 1$).
To remove it, an ionization energy $ \omega_{l}^{0} = -\varepsilon_d $ is needed.
The corresponding spectroscopic 
factors are $f_{u}^{0} = 1$ and $f_{l}^{0} = 1/2$. 
\par
Basically, the exact results presented in Figs.\ \ref{fig_IP}a and \ref{fig_IP}b 
differ from those of this limiting case in two ways. First, in the mixed--valence 
regimes $\varepsilon_d \approx \varepsilon_F$ and $\varepsilon_d \approx \varepsilon_F - U$
the finite coupling causes a smearing effect 
similar to that already observed on the curve $n_d(\varepsilon_d)$ in Fig.\ 
\ref{fig_n_d_G}. Second, the exact curves exhibit numerous avoided crossings. 
In the present case, avoided crossings occur around the points where elementary 
ionization processes become quasi--resonant.
All nearly linear pieces of the curves in Fig.\ \ref{fig_IP}a are reminiscent 
of elementary ionization processes in the noninteracting case ($t_d \to 0$).
For instance, one--hole processes in electrodes are represented by horizontal 
lines. Two--hole-one--particle ($2h$-$1p$) processes where the creation 
of one hole in the upper or lower Hubbard band is accompanied by the excitation 
of a particle--hole pair in electrodes are represented by lines parallel 
to the $\omega_{u}^{0}$-- or $\omega_{l}^{0}$--lines, respectively.
To mention only one more example, 
$2h$-$1p$ processes involving the creation of two holes in the upper and 
lower Hubbard bands 
and of one electron in electrodes at or above the Fermi level
are represented by 
the straight line $\omega_{l\prime}^{0} = - 2 \varepsilon_d - U + \varepsilon_F$ 
or parallels to it. 
At the intersection points of such lines  
avoided crossings arise rather than true intersections, the latter 
occurring only for $t_d \to 0$.
As discussed recently \cite{Baldea:2008}, avoided crossings are very frequent 
in tunable QD systems.
\par
Numerous avoided crossings are visible in Fig.\ \ref{fig_IP}a, but for practical  
purposes they are less important,
and therefore only a few are shown in Fig.\ \ref{fig_IP}b.
Parts thereof are not important because the spectroscopic 
factors are too small. But even if the signals are 
sufficiently intense, avoided crossings are 
less important: in the narrow regions around the critical points where the 
intensities are significant, the energy differences are usually smaller 
than the experimental resolution, and only the summed intensities 
of the participating states can be measured. 
What one measures is a smoothly varying spectroscopic factor, attributed  
to a bright ``diabatic'' state \cite{Baldea:2008}.
The constant spacings between different families of parallel lines in Fig.\ \ref{fig_IP}a 
can be understood by noting that 
the single-particle energies of the isolated electrodes considered there are 
$\varepsilon_F $, $\varepsilon_F \pm t$, and $\varepsilon_F \pm t \sqrt{3} $.
For long electrodes, bundles of dense parallel lines within a width $4 t$ will 
appear, 
so of practical interest are just such bright diabatic states passing through 
a multitude of avoided crossings.
\par
Therefore, besides the results for individual eigenstates, we also show in 
Figs.\ \ref{fig_IP}a and \ref{fig_IP}b the curves $\Omega_{u,l}$ and $f_{u,l}$ 
for the two bright diabatic 
states that are significant. Deep inside the plateaus, the 
ionization signals corresponding to these two diabatic states behave similarly 
to the case $t_d \to 0$, but substantial deviations are visible in the mixed 
valence regimes. Noteworthy is not only that the spectroscopic factors penetrate 
the neighboring charge plateaus, where they gradually decrease,  
similar to $n_d$ (Fig.\ \ref{fig_n_d_G}). Even more interesting is 
the fact that all these penetrations are accompanied by substantial changes 
in the slope of the $\Omega_{u,l}$--curves.
\section{Proposed experiments}
\label{sec-exp} 
In view of the present theoretical results, we propose to conduct the following 
experiments using an intense incoming flux of monochromatic photons, which is well 
focused on the QD of a SET.
To avoid confusions we emphasize again that the photons should have energies of the 
order of the work functions, 
significantly higher than of the microwave and rf--fields considered in the 
previous studies on SETs and the Anderson model.
\par
(i) The first type of experiments is a standard ionization study. 
To measure the absorption coefficient would be 
desirable but presumably a too hard task.  
Probably it would be easier to measure the zero kinetic energy (ZEKE) photoelectrons 
emitted from the dot at 
the ionization threshold, as in threshold ionization experiments in molecular physics.
Just because the QD is very small, not only photoelectrons from the QD, but also from 
electrodes will be inherently ejected into vacuum. Therefore, 
it is important but fortunately easy to experimentally 
distinguish between the ionization of the QD and that of the electrodes:
by varying the gate potential $V_g$, the ionization 
energies $\omega_i$ change in the former case [$\omega_i \approx -\varepsilon_d(V_g)$ and  
$\omega_i \approx -\varepsilon_d(V_g) - U$] 
but remain constant in the latter ($\omega_i = -\varepsilon_F$). 
Let us analyze the change in ionization by gradually increasing 
the dot energy $\varepsilon_d$ starting from a sufficiently low value,
by making $V_g$ more and more negative. 
In the first stage, an ionization signal will be observed, which is characterized by a nearly 
constant intensity $I_u$ ($I_u \propto \tilde{w}_u \simeq 1$). Its intensity will drop
to zero within a narrow range $\delta V_{g 1} \sim 2\Delta V$ ($\Delta V = -\Gamma/\alpha > 0$) 
centered on $V_g = V_{g 1}$.
A little before this signal disappears, another ionization signal 
will appear at $V_g \sim V_{g 1} + \Delta V$, whose intensity rapidly rises to  
a value $I_l \approx I_u/2$ ($I_l \propto \tilde{w}_l \simeq 0.5$) 
beyond $V_g \sim V_{g 1} - \Delta V$. Further on, this intensity remains nearly constant
up to a point where it rapidly falls down to zero within a range 
$\delta V_{g 2} \sim 2\Delta V$ centered on $V_g = V_{g 2}$. 
Except for the (mixed--valence) ranges $\delta V_{g 1}$ and $\delta V_{g 2}$, 
the corresponding ionization 
energies, $\Omega_u(V_g)$ and $\Omega_l(V_g)$, are straight lines of the same slope. 
This slope is half the slope of the $\Omega_l(V_g)$--curve in the range 
$ V_{g 1} < V_g < V_{g 1} + \Delta V$. At the other end, $ V_{g 2} - \Delta V < V_g < V_{g 2}$, 
this curve tends to saturate, $\Omega_l(V_g) \to -\varepsilon_F $. The same tendency 
of the other curve ($\Omega_u(V_g) \to -\varepsilon_F $) can be seen in the opposite 
mixed--valence range $ V_{g 1} - \Delta V < V_g < V_{g 1}$. 
By extrapolating the linear parallel portions of the $\Omega_{u, l}$--curves 
($\Omega_{u,l} \to \tilde{\Omega}_{u,l}$), 
one can \emph{directly} determine the 
charging energy from the difference of their ordinates taken at the same value of 
$V_g$, $U = \tilde{\Omega}_{u}(V_g) - \tilde{\Omega}_{l}(V_g)$ \cite{ZEKE}. 
Noteworthy, for this, 
the values of $V_g$ or $\varepsilon_d$ themselves are \emph{not} needed.
This represents an important difference from the determination of $U$ in 
transport experiments. In the latter, what one can directly measure are the values 
$V_{g 1}$ and $V_{g 2}$, from the positions where the Coulomb blockade peaks occur 
at temperatures $T \agt T_K$ or from the extension of the Kondo plateau observable for  
 $T < T_K$. The determination of $U$ from the difference $V_{g 1} - V_{g 2}$ 
\emph{alone} is impossible, and to this aim the relationship between 
$V_g$ and $\varepsilon_d$, Eq.\ (\ref{eq-epsilon_g}), is necessary.
There is still another important 
difference between ionization and transport measurements. As visible in Fig.\ \ref{fig_IP}a, 
$\Omega_{u}(V_{g 1}) \simeq \Omega_{l}(V_{g 2}) \simeq -\varepsilon_F$. 
The Fermi energy can thus be estimated. This is also important 
because this quantity can alternatively be deduced from the ionization energy 
 $-\varepsilon_F $ of the electrodes (electrode work function). 
To conclude, the $V_g$--dependent  
ionization spectrum allows one to give direct evidence on the formation 
of the charge plateaus and 
to determine the quantities $U$, $\Gamma$, and $\varepsilon_F$. The dependence 
$\varepsilon_d(V_g)$ can also be obtained, which is also important, as it 
allows to quantitatively investigate the capacitances of nanosystems.
\par
(ii) A (ZEKE) ionization study might be difficult, 
but a mixed ionization--transport experiment as proposed below is 
presumably easier while still relevant. In a usual transport experiment, one can measure 
the conductance as a function of the gate potential, $G(V_g)$, 
and determine $V_{g 1}$ and $V_{g 2}$, say, from 
the Kondo plateau ($ T < T_K$). 
At a fixed gate potential $V_{g 2} < V_g < V_{g 1}$ one can now 
shine the QD with radiation whose frequency $\omega$ is varied. At a certain 
value of the latter, $\omega = \omega_{c}(V_g)$, the incoming 
photon will ionize the QD. Because the single (unpaired) electron of the QD 
will be ejected, the Kondo effect will be suppressed, and this will be 
evidenced by the drop in the conductance ($G \approx 0$). 
From the above considerations one expects a linear $V_g$-dependence of 
$\omega_{c}(V_g) \equiv\Omega_{l}(V_g)$. Again, the charging energy can 
be \emph{directly} determined as $U = \tilde{\Omega}_{l}(V_{g 1}) - \tilde{\Omega}_{l}(V_{g 2})$. 
The dependence of $\varepsilon_d$ on $V_g$ is not needed, rather it can be 
deduced by this method, as well as the Fermi energy 
$\varepsilon_F = - \tilde{\Omega}_{l}(V_{g 2})$.
\par
(iii)
A similar mixed experiment can be imagined at $ V_g < V_{g 2}$, 
where the nonionized QD is doubly occupied. At the resonant photon frequency 
$\omega = \Omega_{u}(V_g)$ one electron will be removed. For a sufficiently 
intense irradiation, one can reach a stationary state where the ionized QD 
will be occupied by an unpaired electron. Because the latter is a prerequisite 
for the occurrence of the Kondo effect, it is tempting to 
speculate on a possible photoionization--induced nonequilibrium Kondo 
effect; thence, an effect of photoionization opposite to that of 
suppressing the equilibrium Kondo effect discussed above.
\section{Conclusion}
\label{sec-conclusion} 
In this paper, we have showed that photoionization is a valuable 
method, which can be used along with transport measurements, 
to study SETs. To our knowledge, conductance measurements represent 
the only way to determine the charging energy $U$, which thus resembles 
very much a fit parameter. Therefore, the comparison with the result of 
a completely different 
method is quite meaningful, even more if $U$ is obtained 
in a direct manner, and not via a supplementary relation between 
$\varepsilon_d$ and $V_g$. For specific purposes, photoionization turns 
out to provide even richer information than transport studies. 
\par
The above considerations refer to a QD with a single level. 
However, from the physical analysis 
it is clear that the results are also relevant 
for more general situations. 
\par
Another important result is that QD--ionization can be 
accurately described by considering short electrodes (weak finite-size 
effects), and only very few ionized states are relevant. 
Both aspects are very important, since they urge 
to carry out realistic \emph{ab initio} calculations on photoionization, 
particularly for single-molecule based SETs.
\section*{Acknowledgments}
The authors acknowledge with thanks the financial support for this work 
provided by the Deu\-tsche For\-schungs\-ge\-mein\-schaft (DFG).

\begin{thebibliography}{10}
\bibitem{ZEKE}
Because typically $U \sim 1$\,meV and $\tilde{\Omega}_{u,l} \sim 1$\,eV, it is
  advantageous to use ZEKE--spectroscopy, which has a better resolution than
  standard photoemission.
\end{thebibliography}

\end{document}